\documentclass[prx,twocolumn,floatfix,superscriptaddress,longbibliography,notitlepage]{revtex4-2}

\usepackage{changes}
\setdeletedmarkup{\color{blue}\sout{#1}}
\usepackage{cancel}

\usepackage[many]{tcolorbox}

\newtcolorbox{BoxStyle}{
    sharpish corners, 
    boxrule = 0pt,
    toprule = 4.5pt, 
    enhanced,
    fuzzy shadow = {0pt}{-2pt}{-0.5pt}{0.5pt}{black!35} 
}

\usepackage{graphicx, color, graphpap}

\usepackage{mathrsfs} 
\usepackage{enumitem}
\usepackage{amssymb}
\usepackage{amsthm}
\usepackage{multirow}
\usepackage[colorlinks=true,citecolor=blue,linkcolor=blue,urlcolor=blue]{hyperref}
\usepackage[T1]{fontenc}

\usepackage{bbm}
\usepackage{thmtools,thm-restate}
\usepackage{verbatim}
\usepackage{mathtools}
\usepackage{titlesec}
\usepackage{amsmath}
\usepackage[title]{appendix}
\usepackage[linesnumbered,ruled,vlined]{algorithm2e}
\SetKwInput{kwInit}{Init}

\usepackage[caption = false]{subfig}

\long\def\ca#1\cb{} 

\newcommand{\abs}[2][]{#1| #2 #1|}

\newcommand{\bramatket}[3]{\langle #1 \hspace{1pt} | #2 | \hspace{1pt} #3 \rangle}

\newcommand{\norm}[2][]{#1| \! #1| #2 #1| \! #1|}

\newcommand{\ket}[1]{|#1\rangle}               
\newcommand{\bra}[1]{\langle #1|}              
\newcommand{\dya}[1]{\ket{#1}\!\bra{#1}}

\newcommand{\rank}{\text{rank}}

\newcommand{\Tr}[1]{\mathrm{tr}\left\{#1\right\}}

\newcommand{\Var}{{\text{Var}}}

\newcommand{\ave}[1]{\langle #1\rangle}               
\renewcommand{\geq}{\geqslant}
\renewcommand{\leq}{\leqslant}

\newcommand{\ad}{^\dagger}

\newcommand*{\id}{\openone}

\newcommand{\eq}{\text{eq}}

{}
{}

\usepackage{amssymb}
\usepackage{dsfont}

\begin{document}

\title{{Conditional quantum thermometry -- enhancing precision by measuring less}}

\author{Akira Sone}
\affiliation{Department of Physics, University of Massachusetts, Boston, Massachusetts 02125, USA}
\email{akirasone628@gmail.com}

\author{Diogo O. Soares-Pinto}
\affiliation{Instituto de F\'isica de S\~{a}o Carlos, Universidade de S\~{a}o Paulo, CP 369, 13560-970, S\~{a}o Carlos, S\~{a}o Paulo, Brazil}

\author{Sebastian Deffner}
\affiliation{Department of Physics, University of Maryland, Baltimore County, Baltimore, Maryland 21250, USA}

\begin{abstract}
{Taking accurate measurements of the temperature of quantum systems is a challenging task. The mathematical peculiarities of quantum information make it virtually impossible to measure with infinite precision. In the present paper, we introduce a generalize thermal state, which is conditioned on the pointer states of the available measurement apparatus. We show that this conditional thermal state outperforms the Gibbs state in quantum thermometry. The origin for the enhanced precision can be sought in its asymmetry quantified by the Wigner-Yanase-Dyson skew information. This additional resource is further clarified in a fully resource-theoretic analysis, and we show that there is a Gibbs-preserving map to convert a target state into the conditional thermal state. We relate the quantum J-divergence between the conditional thermal state and the same target state to quantum heat.}
\end{abstract}

\maketitle

\section{Introduction}
Quantum metrology is a task to utilize the peculiar properties of {quantum} systems, such as quantum coherence and {entanglement}, to achieve parameter estimation {at} precision beyond the classical limit~\cite{giovannetti2004,giovannetti2006,giovannetti2011advances,Degen16x}. {Hence, the quest to identify quantum states that are uniquely suited for metrological tasks with limited resources ~\cite{rubio2018non,rubio2019quantum} is of crucial practical relevance}. {In the following, we address this issue and impose the plausible and realistic condition that only the observable \textit{eigenstates} of the measurement apparatus, aka pointer states~\cite{zurek1981pointer, schlosshauer2007decoherence, brasil2015understanding, touil2022eavesdropping,zurek2003decoherence} are available}.

{One of the most prominent metrological tasks is quantum thermometry of stationary states. In fact, quantum thermometry with Gibbs states has been well studied~\cite{correa2015individual,Pasquale16,Sone18a,Sone19a,brunelli2012qubit, mancino2017quantum, campbell2018precision,mehboudi2015thermometry,pasquale2018quantum}}. {In this simplest scenario it is easy to see that} the optimal measurement for estimating the inverse temperature $\beta$ is the Hamiltonian. However, when the system size is very large and {supports quantum} correlations, {even energy measurements are a challenging task}~\cite{Sone18a,Sone19a}. Therefore, {it is desirable to find  \textit{better} states whose corresponding optimal measurements are experimentally implementable,} and which ideally {even} outperform Gibbs states in the low-temperature limit~\cite{mehboudi2019thermometry,narasimhachar2015low,correa2017}. 

\begin{figure}
		\centering
		\includegraphics[width=1\columnwidth]{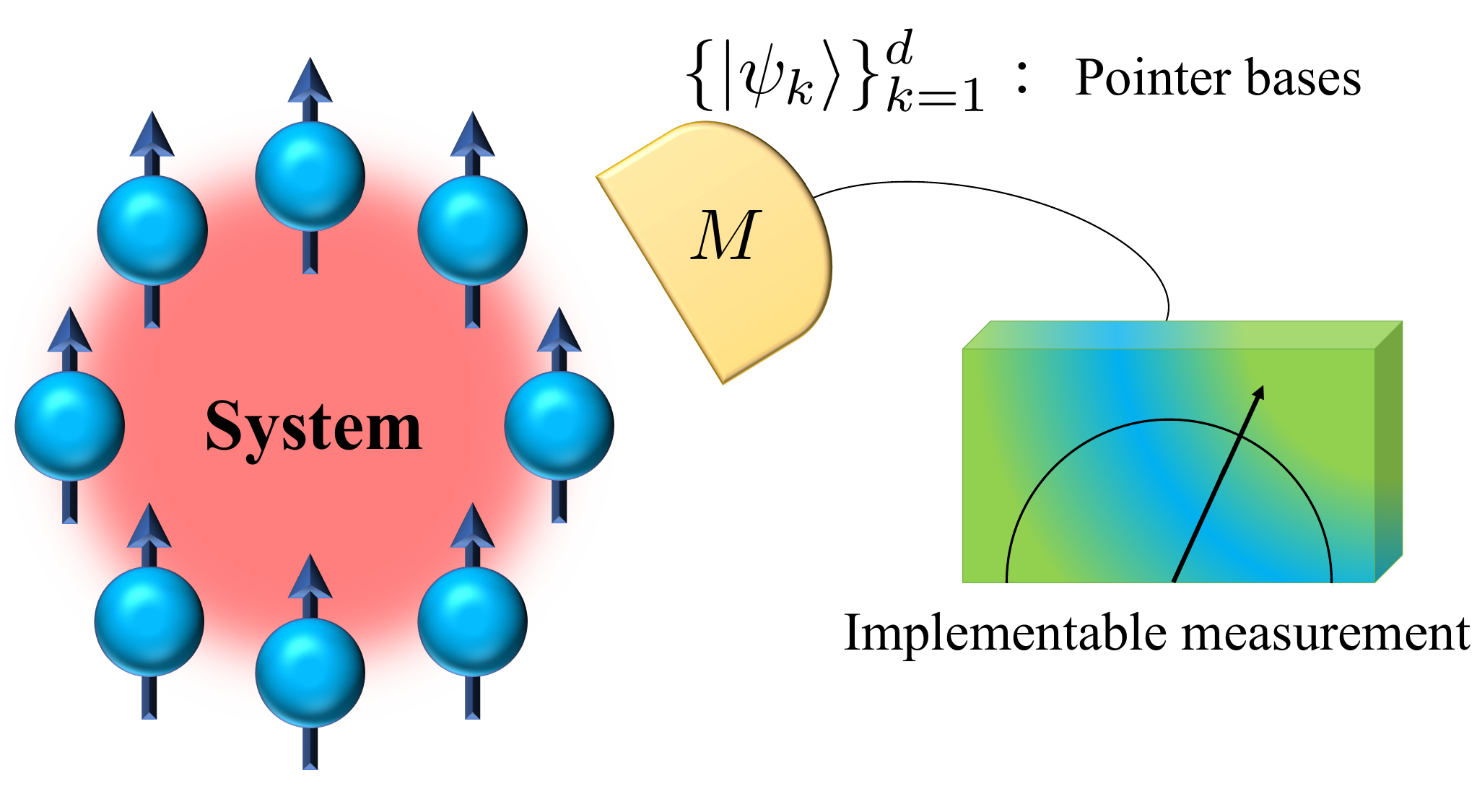}
		\caption{{\textbf{Illustration of the concept}: The \emph{conditional thermal state} (CTS) describes state of a thermal quantum system conditioned on the measurable pointer basis $\{\ket{\psi_k}\}_{k=1}^{d}$ of an implementable measurement $M$. The CTS maximizes the von Neumann entropy under the constraint that the ensemble average of the Hamiltonian is fixed.}}
\label{fig1}
\end{figure}

In this paper, we discuss a quantum state, that indeed fulfills the aforementioned `wishlist'. The \emph{conditional thermal state} (CTS) is constructed as Gibbsian-distributed quantum state of the pointer basis corresponding to the available measurement apparatus. The CTS originally  appeared in the one-time measurement approach to {quantum work}~\cite{Deffner16,Sone20a,Sone21b,Beyer2020,sone2023exchange,sone2023jarzynski} {and correspondingly tighter maximum work theorems. In the following, we} demonstrate that the {CTS outperforms} the Gibbs state in quantum thermometry. {To elucidate its properties further we then show that it can be understood as a non-passive state with useful resources}~\cite{chitambar2019quantum,marvian2014asymmetry,marvian2014asymmetry,gour2008resource,marvian2013theory,gour2009measuring}.  {To this end,} we first relate its asymmetry, which is quantified by the Wigner-Yanase-Dyson (WYD) skew information~\cite{marvian2014extending,takagi2019skew, li2019monotonicity,yamaguchi2023smooth,marvian2016quantum,ahmadi2013wigner}, to its QFI. {Then}, focusing on a system undergoing unitary evolution, we  discuss the state convertibility between the exact final state and the {CTS} constructed by the pointer {states} given by the evolved post-measurement state. Finally, we demonstrate that their symmetric divergence, also known as quantum J-divergence, can be interpreted as quantum heat~\cite{elouard2017role}. {Our} results {demonstrate} the usefulness of the {CTS} as a resource state from both fundamental and practical perspectives.

\section{Conditional thermal state}

{We begin by establishing notions and notations. For the ease of the presentation, and to avoid clutter in the formulas, we work in units for which the Boltzmann constant $k_B$ and the reduced Planck constant $\hbar$ are $k_B=\hbar=1$. 

The concept of the CTS} is illustrated in Fig.~\ref{fig1}. In {a} $d$-dimensional Hilbert space $\mathcal{H}$, given a Hamiltonian $H$ and the pointer {states} $\{\ket{\psi_k}\}_{k=1}^{d}$ of an implementable measurement $M$, the {CTS} is defined as 
\begin{equation}
    \rho_{\beta}\equiv \sum_{k=1}^{d}\frac{e^{-\beta\bramatket{\psi_k}{H}{\psi_k}}}{Z_{\beta}}\dya{\psi_k}\,.
\label{eq:DefConditionalThermal}
\end{equation}
Here, $Z_{\beta}$ is the normalization factor
\begin{equation}
    Z_{\beta}\equiv \sum_{k=1}^{d}e^{-\beta\bramatket{\psi_k}{H}{\psi_k}}\,,
\label{eq:Normalization}
\end{equation}
{which can be interpreted as a generalized partition function~\cite{Deffner16,Sone20a,Sone21b,Beyer2020,sone2023exchange}.} The CTS is defined as the thermal state conditioned on the choice of the measurement implementable in the laboratory. The CTS maximizes the von Neumann entropy~\cite{jaynes1957information} under the constraint that the ensemble average of the Hamiltonian is fixed  {(See Appendix~\ref{app:EntropyMaximization}.)}. Note that when $\{\ket{\psi_k}\}_{k=1}^{d}$ are the pointer {states} of $H$, $\rho_{\beta}$ becomes the Gibbs state $\rho_{\beta}^{\eq}\equiv e^{-\beta H}/Z_{\beta}^{\eq}$, where $Z_{\beta}^{\eq}\equiv\Tr{e^{-\beta H}}$ is the standard canonical partition function. Therefore, {the CTS can be regarded as a generalized thermal state}. 

For {later} convenience, {we} also define the following conditional thermal separable state in a composite Hilbert space $\mathcal{H}_1\otimes\mathcal{H}_2$
\begin{equation}
\label{eq:CTS_sep}
    \widetilde{\rho}_{\beta}\equiv\sum_{k=1}^{d}\frac{e^{-\beta\bramatket{\psi_k}{H}{\psi_k}}}{Z_{\beta}}\dya{\psi_k}_1\otimes\dya{\phi_k}_2\,,
\end{equation}
where $\{\ket{\phi_k}\}_{k=1}^{d}$ are the arbitrary orthogonal {eigenstates} of $\mathcal{H}_2$. From these definitions, we also have $\rho_{\beta}=\mathrm{tr}_2\left\{\widetilde{\rho}_{\beta}\right\}$, 
which {will} become important {in our discussion of} the relation between the {quantum Fisher information (QFI)} for quantum thermometry and the asymmetry measure of $\rho_{\beta}$.

\section{Quantum thermometry}

\subsection{Quantum Fisher information}
An important figure of merit in quantum metrology is {the} QFI. Given a quantum state $\rho_{\theta}$ parameterized by a certain parameter $\theta\in\mathbb{R}$, the variance of an unbiased estimator $(\delta\theta)^2$ is bounded by {the} quantum Cram\'{e}r-Rao bound (QCRB), $(\delta\theta)^2\geq1/\mathcal{I}(\rho_{\theta};\theta)$,
where $\mathcal{I}(\rho_{\theta};\theta)$ is the QFI defined by $\mathcal{I}(\rho_{\theta};\theta)\equiv -2\lim_{\epsilon\to 0}\partial_{\epsilon}^2 \mathcal{F}(\rho_{\theta},\rho_{\theta+\epsilon})
$~\cite{Jing20,fiderer2019maximal,jacobs2014quantum,Sone2020QFI,cerezo2021sub,Beckey2022}. The definition stems from the relation between the Bures distance and the symmetric logarithmic derivative. Thus,  the QFI can be obtained by optimizing the classical Fisher information over all possible measurements, as detailed in Ref.~\cite{jacobs2014quantum}.
Here, $\mathcal{F}(\rho,\sigma)\equiv\norm{\sqrt{\rho}\sqrt{\sigma}}_1^2$ is the quantum fidelity between two quantum states $\rho$ and $\sigma$~\cite{jozsa1994fidelity}, {and} $\norm{A}_1\equiv \Tr{\sqrt{A\ad A}}$ is the trace norm. {We denote $\partial^m/\partial x^m$ simply by $\partial_x^m$}. 

\subsection{Quantum thermometry with the CTS}
{For our present purposes we now analyze the precision, with which the inverse temperature $\beta$ can be estimated from the CTS. This will also allow us to  relate the QFI of the CTS to the WYD skew information $I_{\alpha}(\rho_{\beta},H)$ contained by $\rho_{\beta}$ with respect to the Hamiltonian $H$, which quantifies the asymmetry of $\rho_{\beta}$. }

The QFI $\mathcal{I}(\rho_{\beta};\beta)$ of the {CTS} $\rho_{\beta}$ is given by  {(See Appendix~\ref{app:ProofQFI} for the proof.)}
\begin{equation}
\mathcal{I}(\rho_{\beta};\beta)=\partial_{\beta}^2\ln Z_{\beta}\,, 
\label{eq:CTS-QFI}
\end{equation}
and the optimal measurement achieving the quantum Cram\'{e}r-Rao bound is 
\begin{equation}
G = \sum_{k=1}^{d}\bramatket{\psi_k}{H}{\psi_k} \dya{\psi_k}\,.
\label{eq:OptimalMeasure}
\end{equation}
To compare the sensitivity of the {CTS} $\rho_{\beta}$ and {the} Gibbs state $\rho_{\beta}^{\eq}$, {we define} the QFI difference as $\Delta \mathcal{I}_{\beta}\equiv \mathcal{I}(\rho_{\beta};\beta)-\mathcal{I}(\rho_{\beta}^{\eq};\beta)$. {We have}
\begin{equation}
    \Delta\mathcal{I}_{\beta} = -\partial_{\beta}^2 S(\rho_{\beta}||\rho_{\beta}^{\eq})=\partial_{\beta}^2\left(\ln\frac{Z_{\beta}}{Z_{\beta}^{\eq}}\right)\,,
\label{eq:QFIDifference}
\end{equation}
where $S(\rho_{\beta}||\rho_{\beta}^{\eq})\equiv\Tr{\rho_{\beta}\ln\rho_{\beta}}-\Tr{\rho_{\beta}\ln\rho_{\beta}^{\eq}}$ denotes the quantum relative entropy~\cite{Nielsen} of $\rho_{\beta}$ with respect to $\rho_{\beta}^{\eq}$. {The relative entropy} measures the distinguishability of $\rho_{\beta}$ and $\rho_{\beta}^{\eq}$. Therefore, the condition that the {CTS} outperforms the Gibbs state in quantum thermometry at a certain temperature $\beta_0$ is 
\begin{equation}
\partial_{\beta}^2S(\rho_{\beta}||\rho_{\beta}^{\eq})\Big|_{\beta=\beta_0}<0\,.
\label{eq:OutperformCondition}
\end{equation}
{Therefore,} the curvature of the quantum relative entropy $S(\rho_{\beta}||\rho_{\beta}^{\eq})$ with respect to $\beta$ can be regarded as the criteria quantifying the performance of the {CTS} for quantum thermometry.

\section{Single-qubit example}
\label{sec:singlequbit}
{As a pedagogical example, we show the single-qubit case}. Let $\sigma_x,\sigma_y$ and $\sigma_z$ be the Pauli matrices. The Hamiltonian is $H=\omega \sigma_z$, so that the eigenstates of $H$ are $\ket{0}=\begin{pmatrix}1&0\end{pmatrix}^T$ and $\ket{1}=\begin{pmatrix}0&1\end{pmatrix}^T$. Therefore, $Z_{\beta}^{\eq}=\Tr{e^{-\beta H}}=2\cosh(\beta\omega)$. Considering the pointer {states} $\ket{\psi_0(\theta)}=e^{i\frac{\theta}{2}\sigma_x}\ket{0}$ and $
\ket{\psi_1(\theta)}=e^{i\frac{\theta}{2}\sigma_x}\ket{1}$ with $\theta\in\mathbb{R}$, the normalization factor of the {CTS} becomes
$Z_{\beta}(\theta)=2\cosh(\beta\omega\cos(\theta))$. 

\begin{figure}[htp!]
		\centering
  \includegraphics[width=1.1\columnwidth]{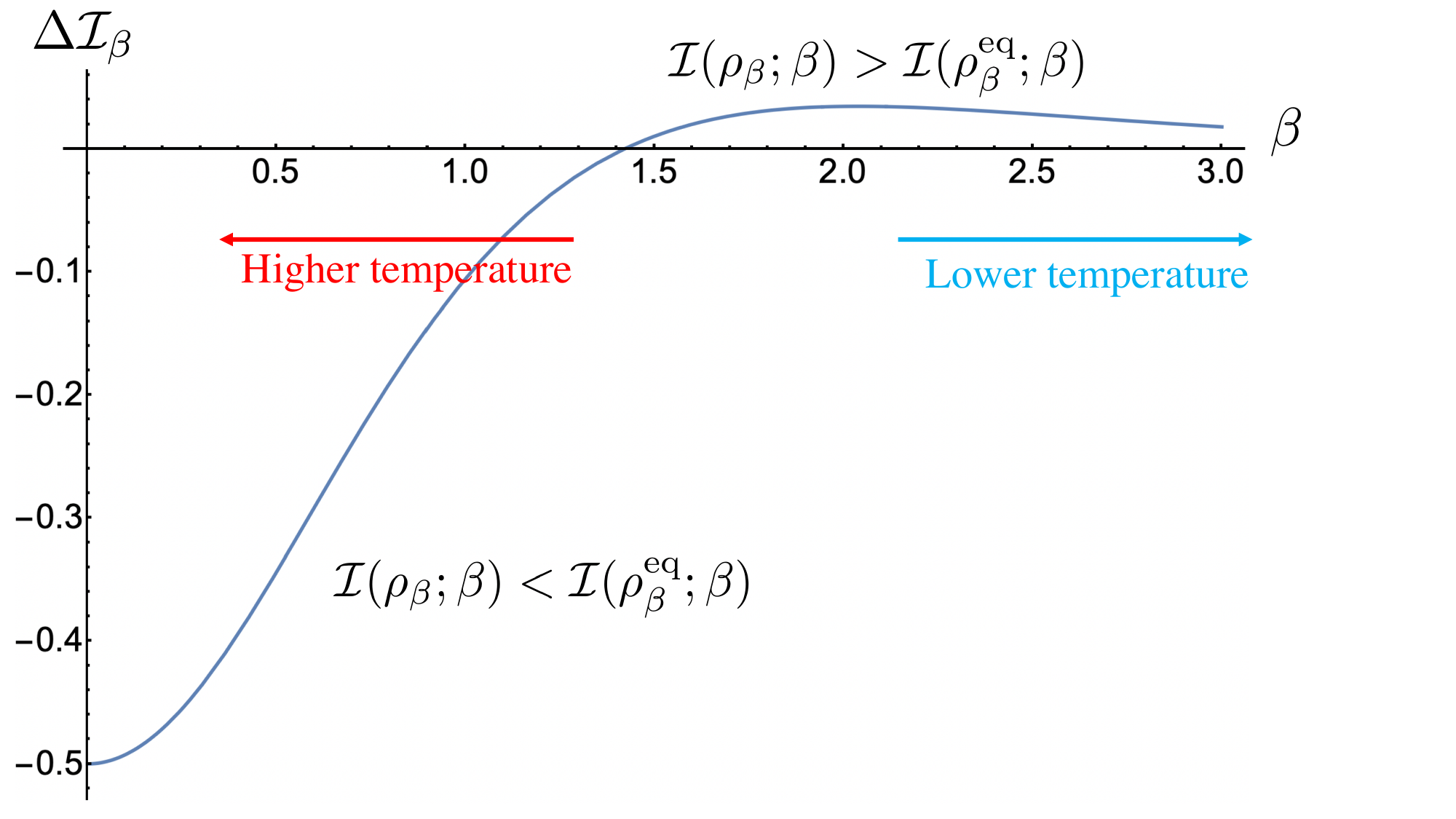}
		\caption{\textbf{Relation between $\Delta I_{\beta}$ and $\beta$}: We choose $\omega=1$ and $\theta=\pi/4$. {In the low-temperature limit,} we have $\mathcal{I}(\rho_{\beta};\beta)\geq \mathcal{I}(\rho_{\beta}^{\eq};\beta)$, meaning that the {CTS} outperforms the Gibbs state in thermometry.}
\label{fig2}
\end{figure}

Therefore, {we obtain from Eq.~\eqref{eq:QFIDifference},}
\begin{equation}
    \Delta\mathcal{I}_{\beta}(\theta)= \omega^2\left(-1+\frac{\cos^2(\theta)}{\cosh^2(\beta\omega\cos(\theta))}+\tanh^2(\beta\omega)\right).
\end{equation}
{In} the low-temperature limit $(\beta\omega\gg1)$, we have 
\begin{equation}
    \Delta\mathcal{I}_{\beta}(\theta)\simeq\left(\frac{\omega\cos(\theta)}{\cosh(\beta\omega\cos(\theta))}\right)^2\geq0 \quad\forall\,\theta\in\mathbb{R}\,.
\end{equation}
{This} means that the {CTS} can outperform the Gibbs state for any {choice of pointer states} in the low-temperature limit. As an example, {choose  $\theta=\pi/4$ and $\omega=1$, which is illustrated in Fig.~\ref{fig2}.}

\section{Asymmetry}

{The natural question arises, where exactly this enhanced performance of the CTS in thermometry originates. To this end, we now analyze} the relation between the QFI of $\rho_{\beta}$ and the asymmetry measure, which is quantified by the WYD skew information $I_{\alpha}(\rho_{\beta},H)$ contained by $\rho_{\beta}$ with respect to the Hamiltonian $H$. 

{The asymmetry of quantum state is related to its quantum coherence, which is regarded as a resource of breaking the symmetry of a group. This is characterized by the \emph{resource theory of asymmetry} (RTA)~\cite{chitambar2019quantum,marvian2014asymmetry,marvian2014asymmetry,gour2008resource,marvian2013theory,gour2009measuring}, whose development has also contributed to 
exploring quantum thermodynamic resources~\cite{gour2015resource,lostaglio2019introductory,lostaglio2015quantum,lostaglio2015description,marvian2022operational,marvian2020coherence}. In particular, when we consider time translations $\{\exp{(-iHt)},|,t\in\mathbb{R}\}$, a state $\rho$ is symmetric if and only if $[\rho,H]=0$, and asymmetric if and only if $[\rho,H]\neq 0$. Here, the symmetric state is considered the free state, while the asymmetric state is regarded as the resource state. Covariant operations~\cite{bartlett2003entanglement}, which are the relevant set of free operations, cannot generate an asymmetric state from a symmetric one or convert one asymmetric state into another. An asymmetry measure must satisfy two conditions: (i) it vanishes if and only if the state is symmetric, and (ii) it does not increase under all covariant operations. One quantity satisfying these conditions is the skew information~\cite{marvian2014extending,takagi2019skew, li2019monotonicity,yamaguchi2023smooth,marvian2016quantum,ahmadi2013wigner}.}

For a quantum state $\rho$, the WYD skew information is defined as $I_{\alpha}(\rho,H)\equiv \Tr{\rho H^2}-\Tr{\rho^{\alpha}H\rho^{1-\alpha}H}$ with $0<\alpha<1$.  With $\Var_{\rho}\{H\}\equiv\Tr{\rho H^2}-\left(\Tr{\rho H}\right)^2$ the variance of $H$ with respect to $\rho$, we can define the general variance (covariance) $Q_{\alpha}(\rho,H)$ as $Q_{\alpha}(\rho, H)\equiv \Var_{\rho}\{H\}-I_{\alpha}(\rho,H)$~\cite{petz2002covariance,manzano2022non,pires2015geometric}.

{Hence, the} QFI of $\rho_{\beta}$ can be alternatively written as the covariance of the Hamiltonian $H_{12}\equiv H\otimes\id_2$ with respect to the conditional thermal separable state $\widetilde{\rho}_{\beta}$ \eqref{eq:CTS_sep}. By using the fact that $\Var_{\widetilde{\rho}_{\beta}}\{H_{12}\}=\Var_{\rho_{\beta}}\{H\}$ and the asymmetry monotone of WYD skew information $I_{\alpha}(\widetilde{\rho}_{\beta},H_{12})\geq I_{\alpha}(\rho_{\beta},H)$~\cite{marvian2014extending,takagi2019skew, li2019monotonicity}, we obtain (See Appendix~\ref{app:QFIvsSkewInfo} for the proof.)
\begin{equation}
\mathcal{I}(\rho_{\beta};\beta) = Q_{\alpha}(\widetilde{\rho}_{\beta},H_{12})\leq Q_{\alpha}(\rho_{\beta},H)\,.
\label{eq:QFIvsSkewInfo}
\end{equation}
Note that when the {chosen basis comprises the pointer states} of the Hamiltonian, i.e., $\rho_{\beta}=\rho_{\beta}^{\eq}$, we have $I_{\alpha}(\rho_{\beta}^{\eq},H)=0$ because of $[\rho_{\beta}^{\eq},H]=0$. In this case, the upper bound becomes $\Var_{\rho_{\beta}}\{H\}$, which is exactly the QFI of $\rho_{\beta}$ for estimating $\beta$, and Eq.~\eqref{eq:QFIvsSkewInfo} saturates. 

These results demonstrate that the asymmetry of the {CTS} quantifies the maximum ultimate precision limit of the quantum thermometry. In our scenario, since the asymmetry resource contains the quantum coherence over the {eigenstates} of the Hamiltonian~\cite{marvian2014extending}, from Eq.~\eqref{eq:QFIvsSkewInfo}, the quantum coherence of the conditional thermal separable state $\widetilde{\rho}_{\beta}$ can be engineered by choosing appropriate pointer {states maximizing} the QFI of $\rho_{\beta}$.

\section{State preparation of the CTS}

{In the preceding section we have shown that the inherent asymmetry of the CTS can be exploited as a resource. Hence, we now continue with a more formal analysis of the resource-theoretic properties of the CTS arising from the state distinguishability with the Gibbs state. To this end, we consider a quantum} system, which is initially prepared in {a Gibbs state, $\rho_{\beta}^{\eq}(0)$, and evolves under a time-dependent Hamiltonian $H_t$ from $t=0$ to $t=\tau$ with a corresponding unitary $U_{\tau}$. Our main interest is} the state convertibility of the exact final state into the {CTS} by considering the information-theoretic properties of the averaged excess work. 

Let us write $\rho(\tau)$ as the exact final state $\rho(\tau) = U_{\tau}\rho_{\beta}^{\eq}(0)U_{\tau}$.
In this case, the exact averaged work $\ave{W}$ is given by $\ave{W}=\Tr{\rho(\tau)H_{\tau}}-\Tr{\rho_{\beta}^{\eq}(0)H_0}$.
When $\Delta F^{\eq}$ is the equilibrium free energy difference $\Delta F^{\eq}\equiv- \beta^{-1}\ln(Z_{\beta}^{\eq}(\tau)/Z_{\beta}^{\eq}(0))$, the excess work is given by~\cite{Deffner2010PRL,landi2021irreversible}
\begin{equation}
    \ave{W_{\text{ex}}}\equiv \ave{W}-\Delta F^{\eq}=\beta^{-1}\,S(\rho(\tau)||\rho_{\beta}^{\eq}(\tau))\,.
\label{eq:ExactDissWork}
\end{equation}
We initially measure the system with $H_0\equiv\sum_{k=1}^{d}E_k\dya{E_k}$, where $\{\ket{E_k}\}_{k=1}^{d}$ are the {eigenstates} of $H_0$. The post-measurement state of the system after the evolution is $U_{\tau}\ket{E_k}$. Thus, the {CTS constructed from} the post-measurement state for the final Hamiltonian $H_{\tau}\equiv\sum_{k=1}^{d}E_k'\dya{E_k'}$ is given by~\cite{Deffner16}
\begin{equation}
    \rho_{\beta}(\tau) = \sum_{k=1}^{d}\frac{e^{-\beta\bramatket{E_k}{U_{\tau}\ad H_{\tau}U_{\tau}}{E_k}}}{Z_{\beta}(\tau)}U_{\tau}\dya{E_k}U_{\tau}\ad\,,
\label{eq:CTS}
\end{equation}
where $Z_{\beta}(\tau)$ is again the {conditional partition function}. 

{In Ref.~\cite{Deffner16} it was shown that} the excess work is lower bounded by
\begin{equation}
    \ave{W_{\text{ex}}}\geq \beta^{-1} S(\rho_{\beta}(\tau)||\rho_{\beta}^{\eq}(\tau))\,.
\label{eq:DissWorkLowerBound}
\end{equation}
Therefore, from Eqs.~\eqref{eq:ExactDissWork} and \eqref{eq:DissWorkLowerBound}, we have
\begin{equation}
    S(\rho(\tau)||\rho_{\beta}^{\eq}(\tau))\geq S(\rho_{\beta}(\tau)||\rho_{\beta}^{\eq}(\tau))\,.
\label{eq:RelativeEntropyInequality}
\end{equation}
 From Refs.~\cite{shiraishi2021quantum,shiraishi2021quantumErratum,takagi2022correlation}, Eq.~\eqref{eq:RelativeEntropyInequality} becomes tight if and only if there exists a sufficiently large $n_0$ and there exists a Gibbs-preserving map $\mathcal{E}_n:\mathcal{H}^{\otimes n}\to\mathcal{H}^{\otimes n}$ for $n\geq n_0$ such  that 
\begin{equation}
\mathcal{E}_n(\rho_{\beta}^{\eq}(\tau)^{\otimes n}) =\rho_{\beta}^{\eq}(\tau)^{\otimes n}\,,~
\mathcal{E}_n(\rho(\tau)^{\otimes n}) =\Xi_n\,,
\label{eq:AsymptoticGibbs}
\end{equation}
where for any $\epsilon>0$, the quantum state $\Xi_n$ satisfies
\begin{equation}
T_n\equiv \frac{1}{2}\norm{\rho_{\beta}(\tau)^{\otimes n}-\Xi_n}_1<\epsilon\,.    
\end{equation}
This means that the CTS is not just a mathematical object but can be prepared, allowing for an arbitrarily small error via the Gibbs-preserving map with an \textit{asymptotic} number of copies of the state.

Also, from the quantum relative entropy of the exact final state $\rho(\tau)$ with respect to the CTS $\rho_{\beta}(\tau)$ $S(\rho(\tau)||\rho_{\beta}(\tau))$, we obtain  {(See Appendix~\ref{app:ThermoTriangleEq} for the proof.)}
\begin{equation}
S(\rho(\tau)||\rho_{\beta}(\tau))\! +\!S(\rho_{\beta}(\tau)||\rho_{\beta}^{\eq}(\tau))  
=S(\rho(\tau)||\rho_{\beta}^{\eq}(\tau))\,,
\label{eq:ThermoTriangleEq}
\end{equation}
which we call \textit{thermodynamic triangle equality}. Therefore, when $S(\rho(\tau)||\rho_{\beta}^{\eq}(\tau))=S(\rho_{\beta}(\tau)||\rho_{\beta}^{\eq}(\tau))$, we must have $\rho(\tau)=\rho_{\beta}(\tau)$ (i.e. $U_{\tau}\ad H_{\tau} U_{\tau}=H_0$). 

To analyze the scaling of the required number of copies $n$ given an arbitrarily small error $\epsilon$, we employ the inequality $S(\rho_{\beta}(\tau)^{\otimes n}||\Xi_n)\leq (T_n+\lambda_n)\ln\left(1+T_n/\lambda_n\right)$~\cite{audenaert2005continuity}, where $\lambda_n$ denotes the minimum non-zero eigenvalue of $\Xi_n$. Since $d=\rank(\rho_{\beta}(\tau))$, we can write $\lambda_n\sim 1/d^n$~\cite{audenaert2005continuity}. Therefore, when $T_n<\epsilon\ll 1$, we have 
\begin{equation}
\begin{split}
S(\rho_{\beta}(\tau)^{\otimes n}||\Xi_n)&\lesssim \left(\epsilon+\frac{1}{d^n}\right)\ln\left(1+\epsilon d^n\right)\\
&=\epsilon\left(1+\sum_{k=1}^{\infty}\frac{(-1)^{k+1}}{k(k+1)}\epsilon^{k}d^{kn}\right)\\
&=\epsilon\left(1 +\frac{d^{n}}{2}\epsilon+O(\epsilon^2 d^{2n})\right)\,.
\end{split}
\end{equation}
To guarantee $\lim_{n\to 0}S(\rho_{\beta}(\tau)^{\otimes n}||\Xi_n)=0$, the simplest condition is $\epsilon d^n\sim 1$. Therefore, the required number of copies of $\rho_{\beta}(\tau)$ can be approximately given by 
\begin{align}
    n\sim \log_{d}\left(\frac{1}{\epsilon}\right)\,.
\end{align}
Here, we note that this state preparation protocol inherently depends on an unknown parameter. Therefore, in the context of thermometry, this protocol remains limited. We need to further explore state preparation protocols that are independent of such parameters.

\section{Quantum heat from the CTS}
Furthermore, the symmetric divergence between $\rho(\tau)$ and $\rho_{\beta}(\tau)$, {i.e., the} quantum J-divergence defined as~\cite{audenaert2013asymmetry}, 
\begin{equation}
J(\rho(\tau),\rho_{\beta}(\tau))\equiv S(\rho(\tau)||\rho_{\beta}(\tau))+S(\rho_{\beta}(\tau)||\rho(\tau))\,,
\end{equation}
is related to the concept of quantum heat~\cite{elouard2017role}. This becomes obvious when the quantum J-divergence is written as  {(See Appendix~\ref{app:Jdivergence} for the proof.)}
\begin{equation}
J(\rho(\tau),\rho_{\beta}(\tau))= \beta\left(\ave{W} - \mathcal{W}_0(\rho_{\beta}(\tau)) - \Delta E(\rho_{\beta}(\tau))\right).
\label{eq:Jdivergence}
\end{equation}
Here, the internal energy change of $\rho_{\beta}(\tau)$ is $\Delta E(\rho_{\beta}(\tau))\equiv \Tr{\rho_{\beta}(\tau)\left(H_{\tau}-H_0\right)}$, and the quantum ergotropy of $\rho_{\beta}(\tau)$ with respect to $H_0$ is  $\mathcal{W}_0(\rho_{\beta}(\tau)) \equiv \Tr{\rho_{\beta}(\tau)H_0}-\Tr{\Gamma\rho_{\beta}(\tau)\Gamma\ad H_0}$, where $\Gamma\equiv\sum_{k=1}^{d}\dya{E_k}U_{\tau}\ad=U_{\tau}\ad$ is the ergotropic transformation~\cite{Allahverdyan04}. From the first law of thermodynamics,  $\beta^{-1}J(\rho(\tau),\rho_{\beta}(\tau))$ can be regarded as a heat, particularly {the} \textit{quantum heat}, which {has been discussed in the literature} as the heat induced by the measurement~\cite{elouard2017role}. In essence, the {CTS} $\rho_{\beta}(\tau)$ is conditioned on the first energy measurement outcome; therefore, its relation to the quantum heat is consistent in this context. 
Also, note that when $U_{\tau}$ is a adiabatic passage, then $\rho(\tau)=\rho_{\beta}(\tau)=\rho_{\beta}^{\eq}(\tau)$, so that $\beta^{-1}J(\rho(\tau),\rho_{\beta}(\tau))=0$, which is consistent {with zero heat exchange} in the adiabatic process.

\section{Conclusion}
In conclusion, we have introduced a conditional thermal state, which is a thermal state conditioned on the pointer {states, and} demonstrated that {this} conditional thermal state can outperform the Gibbs state in the quantum thermometry as a useful resource state. We also have explored its resource-theoretic properties in terms of the asymmetry, state convertibility and its relation to the quantum heat. As an application, these results could help experimentalists to achieve a better sensitivity in quantum thermometry under the constraints in the implementable measurements. From the fundamental point of view, these results provide insightful perspective on the implications of the conditional thermal state in the thermodynamic protocols in the quantum systems in a resource-theoretic approach. {Finally, the present analysis also provides an additional \emph{a posteriori} motivation and justification for the one-time energy measurement approach to quantum work \cite{Deffner16,Sone20a,Sone21b,Beyer2020,sone2023exchange,sone2023jarzynski}}. Future work will involve further exploration of concrete state preparation protocols for generating the conditional thermal state.

\acknowledgements{We would like to thank to Ryuji Takagi for helpful discussions. AS gratefully acknowledges startup funding supported by the University of Massachusetts, Boston. DOSP acknowledges the Brazilian funding agencies CNPq (Grant No. 307028/2019-4), FAPESP (Grant No. 2017/03727-0), and the Brazilian National Institute of Science and Technology of Quantum Information (INCT-IQ) Grant No. 465469/2014-0. {S.D. acknowledges support from the John Templeton Foundation under Grant No. 62422.}}

\bibliography{ref.bib}

\onecolumngrid
	
	\setcounter{section}{0}
	\setcounter{theorem}{0}
	\setcounter{lemma}{0}
	\setcounter{definition}{0}

\appendix

\section{Derivation of conditional thermal state}
\label{app:EntropyMaximization}
Consider a density matrix $\rho$ with the following spectral decomposition
\begin{equation}
    \rho = \sum_{k=1}^{d}p_k\dya{\psi_k}\,,
\end{equation}
where $\{\ket{\psi_k}\}_{k=1}^{d}$ are the pointer {states} of a given measurement $M$. Let $H$ be the Hamiltonian of the system. Then, we want to maximize the von Neumann entropy
\begin{equation}
S(\rho)\equiv -\Tr{\rho\ln\rho}
\label{eq:defEntropy}
\end{equation}
under the conditions
\begin{equation}
\begin{split}
\Tr{\rho}=1\quad\text{and}\quad E(\rho)\equiv\Tr{\rho H}=\text{Const}. 
\end{split}
\label{eq:Constraint}
\end{equation}
By using the optimization method of Lagrange multipliers with constraints, we have: 
\begin{equation}
\delta\left(S(\rho)-\gamma E(\rho)-\alpha\right)\\
=-\sum_{k=1}^{d}\delta p_k\left(\ln p_k+\beta\bramatket{\psi_k}{H}{\psi_k}+\alpha+1\right)
=0\,.
\label{eq:Lagrange}
\end{equation}
For any $\delta p_k$, Eq.~\eqref{eq:Lagrange} has to be valid so that each term has to be independently 0. Therefore, we must have $p_k \propto e^{-\beta\bramatket{\psi_k}{H}{\psi_k}}$. From the requirement of $\sum_{k=1}^{d}p_k=1$, we obtain the {CTS} $\rho_{\beta}$.

\section{Detailed derivations of Eqs.~\eqref{eq:CTS-QFI}, \eqref{eq:OptimalMeasure}, \eqref{eq:QFIDifference} and \eqref{eq:OutperformCondition}}
\label{app:ProofQFI}
From the definition of QFI, we have
\begin{equation}
    \mathcal{I}(\rho_{\theta};\theta)= -2\lim_{\epsilon\to 0}\partial_{\epsilon}^2 \mathcal{F}(\rho_{\theta},\rho_{\theta+\epsilon})
\label{eq:app:QFIDef}
\end{equation}
where 
\begin{equation}
\begin{split}
    \mathcal{F}\left(\rho_{\beta},\rho_{\beta+\epsilon}\right)&\equiv\norm{\sqrt{\rho_{\beta}}\sqrt{\rho_{\beta+\epsilon}}}_{1}^{2}\\
    &\equiv\left( \Tr{\sqrt{\rho_{\beta}^{1/2}\rho_{\beta+\epsilon}\,\rho_{\beta}^{1/2}}}\right)^2\\
    &=\frac{1}{Z_{\beta}Z_{\beta+\epsilon}}\left(\sum_{k=1}^{d}e^{-(\beta+\frac{\epsilon}{2})\bramatket{\psi_k}{H}{\psi_k}}\right)^2\\
    &=\frac{Z_{\beta+\epsilon/2}^2}{Z_{\beta}Z_{\beta+\epsilon}}\,.
\end{split}
\end{equation}

Before calculating the quantum Fisher information, let us show the following fact:
\begin{equation}
\begin{split}
\lim_{\epsilon\to0}\partial_{\epsilon}Z_{\beta+\epsilon}=&-\sum_{k=1}^{d}e^{-\beta\bramatket{\psi_k}{H}{\psi_k}}\bramatket{\psi_k}{H}{\psi_k}=\partial_{\beta}Z_{\beta}\\
\lim_{\epsilon\to0}\partial_{\epsilon}^2Z_{\beta+\epsilon}=&\sum_{k=1}^{d}e^{-\beta\bramatket{\psi_k}{H}{\psi_k}}\bramatket{\psi_k}{H}{\psi_k}^2=\partial_{\beta}^2 Z_{\beta}\\
\lim_{\epsilon\to0}\partial_{\epsilon}Z_{\beta+\epsilon/2}=&-\frac{1}{2}\sum_{k=1}^{d}e^{-\beta\bramatket{\psi_k}{H}{\psi_k}}\bramatket{\psi_k}{H}{\psi_k}=\frac{1}{2}\partial_{\beta}Z_{\beta}\\
\lim_{\epsilon\to0}\partial_{\epsilon}^2Z_{\beta+\epsilon/2}=&\frac{1}{4}\sum_{k=1}^{d}e^{-\beta\bramatket{\psi_k}{H}{\psi_k}}\bramatket{\psi_k}{H}{\psi_k}^2=\frac{1}{4}\partial_{\beta}^2 Z_{\beta}\,.
\end{split}
\end{equation}

For two functions $f(x)$ and $g(x)$, where $g(x)\neq0$, we have:  
\begin{equation*}
\partial_{x}^2\left(\frac{f^2}{g}\right)=2 (\partial_x^2 f)\frac{f}{g}+\frac{2}{g}(\partial_x f)^2-\frac{4f}{g^2}(\partial_x f)(\partial_x g)-(\partial_x^2 g)\frac{f^2}{g^2}+\frac{2 f^2}{g^3}(\partial_x g)^2.
\end{equation*}
Therefore, if we define $x=\epsilon$, $f=Z_{\beta+\epsilon/2}$ and $g=Z_{\beta+\epsilon}$, we can obtain:
\begin{equation}
\begin{split}
\lim_{\epsilon\to0}\partial_{\epsilon}^2 F(\rho_{\beta},\rho_{\beta+\epsilon})&=\lim_{\epsilon\to0}\partial_{\epsilon}^2\left(\frac{Z_{\beta+\epsilon/2}^2}{Z_{\beta}Z_{\beta+\epsilon}}\right)\\
&=\frac{1}{2}\frac{\partial_{\beta}^2Z_{\beta}}{Z_{\beta}} +\frac{1}{2}\left(\frac{\partial_{\beta}Z_{\beta}}{Z_{\beta}}\right)^2-2\left(\frac{\partial_{\beta}Z_{\beta}}{Z_{\beta}}\right)^2-\frac{\partial_{\beta}^2 Z_{\beta}}{Z_{\beta}}+2\left(\frac{\partial_{\beta}Z_{\beta}}{Z_{\beta}}\right)^2\\
&=\frac{1}{2}\left(\frac{\partial_{\beta}Z_{\beta}}{Z_{\beta}}\right)^2-\frac{1}{2}\frac{\partial_{\beta}^2 Z_{\beta}}{Z_{\beta}}\\
&=\frac{1}{2}\left((\partial_{\beta}\ln Z_{\beta})^2-Z_{\beta}^{-1}\partial_{\beta}^2 Z_{\beta}\right)
\end{split}
\end{equation}

Therefore, from Eq.~\eqref{eq:app:QFIDef}, we can obtain
\begin{equation}
    \mathcal{I}(\rho_{\beta};\beta) =  Z_{\beta}^{-1}\partial_{\beta}^2 Z_{\beta}
    -\left(\partial_{\beta}\ln Z_{\beta}\right)^2= \partial_{\beta}^2\ln Z_{\beta}\,.
\label{eq:app:QFI}
\end{equation}

For the optimal measurement, we need to prove that the uncertainty of the unbiased estimator with the measurement 
\begin{equation}
G=\sum_{k=1}^{d}\bramatket{\psi_k}{H}{\psi_k}\dya{\psi_k}
\label{eq:app:OptimalMeasurement}
\end{equation}
achieves the quantum Cram\'{e}r-Rao bound. The uncertainty is given by 
\begin{equation}
    \left(\delta\beta\right)^2_{G} \equiv \frac{\Var_{\rho_{\beta}}\{G\}}{\abs{\partial_{\beta}\Tr{\rho_{\beta}G}}^2}\,.
\label{eq:app:uncertainty}
\end{equation}
Here, we have 
\begin{equation}
    \Tr{\rho_{\beta}G}=\sum_{k=1}^{d}\frac{e^{-\beta\bramatket{\psi_k}{H}{\psi_k}}}{Z_{\beta}}\bramatket{\psi_k}{H}{\psi_k}=\Tr{\rho_{\beta}H}=-\partial_{\beta}\ln Z_{\beta}
\end{equation}
and
\begin{equation}
\begin{split}
\Var_{\rho_{\beta}}\{G\} &\equiv \Tr{\rho_{\beta}H^2}-\left(\Tr{\rho_{\beta}H}\right)^2\\
    &= \sum_{k=1}^{d}\frac{e^{-\beta\bramatket{\psi_k}{H}{\psi_k}}}{Z_{\beta}}\bramatket{\psi_k}{H}{\psi_k}^2-(\partial_{\beta}\ln Z_{\beta})^2\\
&=Z_{\beta}^{-1}\partial_{\beta}^2Z_{\beta}-(\partial_{\beta}\ln Z_{\beta})^2\\
    &=\partial_{\beta}^2\ln Z_{\beta}\,.
\end{split}
\end{equation}
From Eqs.~\eqref{eq:app:QFI} and \eqref{eq:app:uncertainty}, we can obtain 
\begin{equation} 
\left(\delta\beta\right)_{G}^2 =\frac{1}{\partial_{\beta}^2\ln Z_{\beta}}= \frac{1}{\mathcal{I}(\rho_{\beta};\beta)}\,,
\end{equation}
which is the quantum Cram\'{e}r-Rao bound; therefore, Eq.~\eqref{eq:app:OptimalMeasurement} is the optimal measurement for the quantum thermometry.  

From Eq.~\eqref{eq:app:QFI}, we can 
write
\begin{equation}
    \Delta\mathcal{I}_{\beta}\equiv\mathcal{I}(\rho_{\beta};\beta)-\mathcal{I}(\rho_{\beta}^{\eq};\beta) = \partial_{\beta}^2\ln\frac{Z_{\beta}}{Z_{\beta}^{\eq}}\,.
\end{equation}
Since the von Neumann entropy  $S(\rho_{\beta})\equiv-\Tr{\rho_{\beta}\ln\rho_{\beta}}$ is
\begin{equation}
    S(\rho_{\beta}) = \beta\Tr{\rho_{\beta}H}+\ln Z_{\beta}\,,
\end{equation}
the quantum relative entropy $S(\rho_{\beta}||\rho_{\beta}^{\eq})$ becomes
\begin{equation}
    S(\rho_{\beta}||\rho_{\beta}^{\eq}) = -S(\rho_{\beta})-\Tr{\rho_{\beta}\ln\rho_{\beta}^{\eq}}=-\ln\frac{Z_{\beta}}{Z_{\beta}^{\eq}}\,.
\end{equation}
Therefore, we can write
\begin{equation}
    \Delta\mathcal{I}_{\beta} = -\partial_{\beta}^2S(\rho_{\beta}||\rho_{\beta}^{\eq})\,.
\end{equation}
When the {CTS} outperforms the Gibbs state for the quantum thermometry at $\beta=\beta_0$, i.e. 
\begin{equation}
    \Delta\mathcal{I}_{\beta_0} >0\,,
\end{equation}
we must have 
\begin{equation}
    \partial_{\beta}^2S(\rho_{\beta}||\rho_{\beta}^{\eq})\Big|_{\beta=\beta_0}<0\,.
\end{equation}

\section{Proof of Eq.~\eqref{eq:QFIvsSkewInfo}}
\label{app:QFIvsSkewInfo}
First, let us prove $\mathcal{I}(\rho_{\beta};\beta)=Q_{\alpha}(\widetilde{\rho}_{\beta},H_{12})$, where 
\begin{equation}
H_{12}\equiv H\otimes\id_2\,.
\end{equation}
Since
\begin{equation}
    \widetilde{\rho}_{\beta}\equiv\sum_{k=1}^{d} \frac{e^{-\beta\bramatket{\psi_k}{H}{\psi_k}}}{Z_{\beta}}\dya{\psi_k}_1\otimes\dya{\phi_k}_2
\end{equation}
and
\begin{equation}
    \rho_{\beta}=\mathrm{tr}_{2}\{\widetilde{\rho}_{\beta}\}\,,
\end{equation}
we have
\begin{equation}
    \Tr{\widetilde{\rho}_{\beta}H_{12}} = \Tr{\rho_{\beta}H}=-\partial_{\beta}\ln Z_{\beta}
\end{equation}
and
\begin{equation}
    \Tr{\widetilde{\rho}_{\beta}H_{12}^2} = \Tr{\rho_{\beta}H^2}\,,
\end{equation}
so that 
\begin{equation}
    \Var_{\widetilde{\rho}_{\beta}}\{H_{12}\} = \Var_{\rho_{\beta}}\{H\}\,.
\label{eq:app:Variance}
\end{equation}

Next, let us compute the WYD skew information $I_{\alpha}(\widetilde{\rho}_{\beta},H)$ contained by $\widetilde{\rho}_{\beta}$ with respect to $H_{12}$. By using the fact that for a pure state $\ket{\varphi}$ we have $\dya{\varphi}^{\alpha}=\dya{\varphi}$ for any $\alpha$, we can obtain 
\begin{equation}
\begin{split}
    I_{\alpha}(\widetilde{\rho}_{\beta},H_{12}) &= \Tr{\widetilde{\rho}_{\beta}H_{12}^2} -\Tr{\widetilde{\rho}_{\beta}^{\alpha}H_{12}\widetilde{\rho}_{\beta}^{1-\alpha}H_{12}}\\
    &=\Tr{\widetilde{\rho}_{\beta}H_{12}^2} - \sum_{k=1}^{d}\frac{e^{-\beta\bramatket{\psi_k}{H}{\psi_k}}}{Z_{\beta}}\bramatket{\psi_k}{H}{\psi_k}^2\\
    &=\Tr{\widetilde{\rho}_{\beta}H_{12}^2}-Z_{\beta}^{-1}\partial_{\beta}^2 Z_{\beta}\,.
\end{split}    
\end{equation}
Therefore, we can obtain
\begin{equation}
\begin{split}
Q_{\alpha}(\widetilde{\rho}_{\beta},H_{12})
&\equiv
\Var_{\widetilde{\rho}_{\beta}}\{H\}-I_{\alpha}(\widetilde{\rho}_{\beta},H_{12})\\
&=Z_{\beta}^{-1}\partial_{\beta}^2Z_{\beta}-(\partial_{\beta}\ln Z_{\beta})^2\\
 &=\partial_{\beta}^2\ln Z_{\beta}\,.
\end{split}
\end{equation}
From Eq.~\eqref{eq:app:QFI}, we can finally write
\begin{equation}
\mathcal{I}(\rho_{\beta};\beta)=Q_{\alpha}(\widetilde{\rho}_{\beta},H_{12})\,.
\label{eq:app:QFISkew}
\end{equation}
Next, let us prove $I(\rho_{\beta};\beta)\leq Q_{\alpha}(\rho_{\beta},H)$. The WYD skew information has asymmetry monotone~\cite{takagi2019skew}
\begin{equation}
    I_{\alpha}(\widetilde{\rho}_{\beta},H_{12})\geq I_{\alpha}(\rho_{\beta},H)\,.
\end{equation}
From Eq.~\eqref{eq:app:Variance}, we can obtain 
\begin{equation}
    Q_{\alpha}(\widetilde{\rho}_{\beta},H_{12})\leq Q_{\alpha}(\rho_{\beta},H)
    =\Var_{\rho_{\beta}}\{H\}-I_{\alpha}(\rho_{\beta},H)\,.
\end{equation}
Because of Eq.~\eqref{eq:app:QFISkew}, we can obtain
\begin{equation}
\mathcal{I}(\rho_{\beta};\beta)\leq Q_{\alpha}(\rho_{\beta},H)\,.
\end{equation}

\section{Proof of Eq.~\eqref{eq:ThermoTriangleEq}}
\label{app:ThermoTriangleEq}
The quantum relative entropy $S(\rho(\tau)||\rho_{\beta}(\tau))$ can be written as 
\begin{equation}
\begin{split}
S(\rho(\tau)||\rho_{\beta}(\tau))&=-S(\rho(\tau))-\Tr{\rho(\tau)\ln\rho_{\beta}(\tau)}\\
&=-S(\rho(\tau))+\beta\sum_{k=1}^{d}\bramatket{E_k}{U_{\tau}\ad \rho(\tau)U_{\tau}}{E_k}\bramatket{E_k}{U_{\tau}\ad H_{\tau} U_{\tau}}{E_k}+\ln Z_{\beta}(\tau)\,.
\end{split}
\end{equation}
Because 
\begin{equation}
    \rho(\tau) = U_{\tau}\rho_{\beta}^{\eq}(0)U_{\tau}\ad = \sum_{k=1}^{d}\frac{e^{-\beta E_k}}{Z_{\beta}^{\eq}(0)}U_{\tau}\dya{E_k}U_{\tau}\ad\,,
\end{equation}
we have 
\begin{equation}
\sum_{k=1}^{d}\bramatket{E_k}{U_{\tau}\ad\rho(\tau)U_{\tau}}{E_k}\bramatket{E_k}{U_{\tau}\ad H_{\tau}U_{\tau}}{E_k} =\Tr{U_{\tau}\rho_{\beta}^{\eq}(0)U_{\tau}\ad H_{\tau}}= \Tr{\rho(\tau)H_{\tau}}\,.
\end{equation}
Therefore, 
\begin{equation}
    S(\rho(\tau)||\rho_{\beta}(\tau)) = -S(\rho(\tau))+\beta\Tr{\rho(\tau)H_{\tau}}+\ln Z_{\beta}(\tau)\,.
\end{equation}
From the unitary invariance of the von Neumann entropy, we can write
\begin{equation}
\begin{split}
S(\rho(\tau)||\rho_{\beta}(\tau)) & = -S(\rho_{\beta}^{\eq}(0))+\ln Z_{\beta}(\tau)+\beta\Tr{\rho(\tau)H_{\tau}}\\
&=\beta\left(\Tr{\rho(\tau)H_{\tau}}-\Tr{\rho_{\beta}^{\eq}(0)H_0}\right)+\ln\frac{Z_{\beta}(\tau)}{Z_{\beta}^{\eq}(0)}\\
&= \beta \langle W\rangle + \ln\frac{Z_{\beta}(\tau)}{Z_{\beta}^{\eq}(0)}\,,
\end{split}
\end{equation}
where 
\begin{equation}
\langle W\rangle \equiv \Tr{\rho(\tau)H_{\tau}}-\Tr{\rho_{\beta}^{\eq}(0)H_0}
\end{equation}
is the exact averaged work. 
Because
\begin{equation}
    \ln\frac{Z_{\beta}(\tau)}{Z_{\beta}^{\eq}(0)} = \ln\frac{Z_{\beta}(\tau)}{Z_{\beta}^{\eq}(\tau)}+\ln\frac{Z_{\beta}^{\eq}(\tau)}{Z_{\beta}^{\eq}(0)}\,,
\end{equation}
from Ref.~\cite{Deffner16}, we have
\begin{equation}
  \ln\frac{Z_{\beta}(\tau)}{Z_{\beta}^{\eq}(0)} =  -S(\rho_{\beta}(\tau)||\rho_{\beta}^{\eq}(\tau))-\beta\Delta F^{\eq}\,. 
\label{eq:app:Zratio}
\end{equation}
Therefore, we have
\begin{equation}
    S(\rho(\tau)||\rho_{\beta}(\tau))+S(\rho_{\beta}(\tau)||\rho_{\beta}^{\eq}(\tau))=\beta\left(\ave{W}-\Delta F^{\eq}\right)\,.
\label{eq:app:Relative1}
\end{equation}
Because the exact excess work is~\cite{landi2021irreversible}
\begin{equation}
    \ave{W_{\text{ex}}} 
    = \ave{W}-\Delta F^{\eq}
    =\beta^{-1}S(\rho(\tau)||\rho_{\beta}^{\eq}(\tau))\,,
\end{equation}
we can obtain the thermodynamic triangle equality
\begin{equation}
S(\rho(\tau)||\rho_{\beta}(\tau)) +S(\rho_{\beta}(\tau)||\rho_{\beta}^{\eq}(\tau))  
=S(\rho(\tau)||\rho_{\beta}^{\eq}(\tau))\,.
\label{eq:app:ThermoTriangle}
\end{equation}

\section{Proof of Eq.~\eqref{eq:Jdivergence}}
\label{app:Jdivergence}

The quantum J-divergence of $\rho(\tau)$ and $\rho_{\beta}(\tau)$ is defined as 
\begin{equation}
    J(\rho(\tau),\rho_{\beta}(\tau)) \equiv S(\rho(\tau)||\rho_{\beta}(\tau))+S(\rho_{\beta}(\tau)||\rho(\tau))\,.
\label{eq:app:quantumJ}
\end{equation}

In Eqs.~\eqref{eq:app:Relative1} and \eqref{eq:app:ThermoTriangle}, we have already proved the thermodynamic triangle equality. Next, let us consider the quantum relative entropy  $S(\rho_{\beta}(\tau)||\rho(\tau))$, which is given by
\begin{equation}
    S(\rho_{\beta}(\tau)||\rho(\tau))=-S(\rho_{\beta}(\tau))-\Tr{\rho_{\beta}(\tau)\ln\rho(\tau)}\,.
\end{equation}
The von Neumann entropy of $\rho_{\beta}(\tau)$ is 
\begin{equation}
    S(\rho_{\beta}(\tau)) = \beta\Tr{\rho_{\beta}(\tau)H_{\tau}}+\ln Z_{\beta}(\tau)\,.
\end{equation}
Also
\begin{equation}
\begin{split}
    \Tr{\rho_{\beta}(\tau)\ln\rho(\tau)}& =-\beta\sum_{k=1}^{d}E_k \bramatket{E_k}{U_{\tau}\ad\rho_{\beta}(\tau)U_{\tau}}{E_k}-\ln Z_{\beta}^{\eq}(0)\\
    &=-\beta\sum_{k=1}^{d}E_k\frac{e^{-\beta\bramatket{E_k}{U_{\tau}\ad H_{\tau}U_{\tau}}{E_k}}}{Z_{\beta}(\tau)}-\ln Z_{\beta}^{\eq}(0)\,.
\end{split}
\end{equation}
Therefore,
\begin{equation}
    S(\rho_{\beta}(\tau)||\rho(\tau)) =
    \beta\left(\sum_{k=1}^{d}E_k\frac{e^{-\beta\bramatket{E_k}{U_{\tau}\ad H_{\tau}U_{\tau}}{E_k}}}{Z_{\beta}(\tau)}-\Tr{\rho_{\beta}(\tau)H_{\tau}}\right)
    -\ln\frac{Z_{\beta}(\tau)}{Z_{\beta}^{\eq}(0)}\,.
\end{equation}
From Eq.~\eqref{eq:app:Zratio}, we can obtain 
\begin{equation}
    S(\rho_{\beta}(\tau)||\rho(\tau)) - S(\rho_{\beta}(\tau)||\rho_{\beta}^{\eq}(\tau))-\beta \Delta F_{\beta}^{\eq} = \beta\left(\sum_{k=1}^{d}E_k\frac{e^{-\beta\bramatket{E_k}{U_{\tau}\ad H_{\tau}U_{\tau}}{E_k}}}{Z_{\beta}(\tau)}-\Tr{\rho_{\beta}(\tau)H_{\tau}}\right)\,. 
\end{equation}
By using $\Tr{\rho_{\beta}(\tau)H_{0}}$, we can write
\begin{equation}
\begin{split}
    \sum_{k=1}^{d}E_k\frac{e^{-\beta\bramatket{E_k}{U_{\tau}\ad H_{\tau}U_{\tau}}{E_k}}}{Z_{\beta}(\tau)}-\Tr{\rho_{\beta}(\tau)H_{\tau}} =\sum_{k=1}^{d}E_k\frac{e^{-\beta\bramatket{E_k}{U_{\tau}\ad H_{\tau}U_{\tau}}{E_k}}}{Z_{\beta}(\tau)}-\Tr{\rho_{\beta}(\tau)H_0}-\Tr{\rho_{\beta}(\tau)(H_{\tau}-H_0)} 
\end{split}
\end{equation}

Let $\Gamma$ be the ergotropic transformation
\begin{equation}
\Gamma \equiv \sum_{k=1}^{d}\dya{E_k}U_{\tau}\ad=U_{\tau}\ad\,.
\end{equation}
Then, we have
\begin{equation}
    \Tr{\Gamma \rho_{\beta}(\tau)\Gamma\ad H_0} = \sum_{k=1}^{d}E_k\frac{e^{-\beta\bramatket{E_k}{U_{\tau}\ad H_{\tau} U_{\tau}}{E_k}}}{Z_{\beta}(\tau)}\,.
\end{equation}
The ergotropy of $\rho_{\beta}(\tau)$ with respect to the Hamiltonian $H_0$ is defined as~\cite{Allahverdyan04}
\begin{equation}
\mathcal{W}_0(\rho_{\beta}(\tau))\equiv \Tr{\rho_{\beta}(\tau)H_0}-\Tr{\Gamma \rho_{\beta}(\tau)\Gamma\ad H_0}
=\Tr{\rho_{\beta}(\tau)H_0}-\sum_{k=1}^{d}E_k\frac{e^{-\beta\bramatket{E_k}{U_{\tau}\ad H_{\tau} U_{\tau}}{E_k}}}{Z_{\beta}(\tau)}.
\end{equation}
Therefore, by defining 
\begin{equation}
    \Delta E(\rho_{\beta}(\tau))\equiv \Tr{\rho_{\beta}(\tau)(H_{\tau}-H_0)}\,,
\end{equation}
we can write
\begin{equation}
  \sum_{k=1}^{d}E_k\frac{e^{-\beta\bramatket{E_k}{U_{\tau}\ad H_{\tau}U_{\tau}}{E_k}}}{Z_{\beta}(\tau)}-\Tr{\rho_{\beta}(\tau)H_{\tau}}  = -\mathcal{W}_0(\rho_{\beta}(\tau))-\Delta E(\rho_{\beta}(\tau))\,.
\end{equation}
Therefore, 
\begin{equation}
    S(\rho_{\beta}(\tau)||\rho(\tau)) - S(\rho_{\beta}(\tau)||\rho_{\beta}^{\eq}(\tau)) = \beta\left(\Delta F^{\eq}-\mathcal{W}_0(\rho_{\beta}(\tau))-\Delta E(\rho_{\beta}(\tau))\right)\,. 
\label{eq:app:Relative2}
\end{equation}
From Eqs.~\eqref{eq:app:quantumJ}, \eqref{eq:app:Relative1} and \eqref{eq:app:Relative2}, we can obtain
\begin{equation}
    J(\rho(\tau),\rho_{\beta}(\tau))= \beta\left(\ave{W}-\mathcal{W}_0(\rho_{\beta}(\tau))-\Delta E(\rho_{\beta}(\tau))\right)\,.
\end{equation}

\end{document}